\documentclass[letterpaper]{article} % DO NOT CHANGE THIS
\usepackage{aaai25}  % DO NOT CHANGE THIS
\usepackage{times}  % DO NOT CHANGE THIS
\usepackage{helvet}  % DO NOT CHANGE THIS
\usepackage{courier}  % DO NOT CHANGE THIS
\usepackage[hyphens]{url}  % DO NOT CHANGE THIS
\usepackage{graphicx} % DO NOT CHANGE THIS
\usepackage{natbib}  % DO NOT CHANGE THIS AND DO NOT ADD ANY OPTIONS TO IT
\usepackage{caption} % DO NOT CHANGE THIS AND DO NOT ADD ANY OPTIONS TO IT
\usepackage{amssymb}
\usepackage{amsmath}
\usepackage{enumerate}
\usepackage{booktabs}
\usepackage{multirow} 
\usepackage{newfloat}
\usepackage{bbding}
\usepackage{placeins}
\usepackage{float}
\usepackage[table,xcdraw]{xcolor}
\usepackage{algorithm}
\usepackage{algorithmic}

\usepackage{newfloat}
\usepackage{listings}
\DeclareCaptionStyle{ruled}{labelfont=normalfont,labelsep=colon,strut=off} % DO NOT CHANGE THIS
\floatstyle{ruled}
\newfloat{listing}{tb}{lst}{}
\floatname{listing}{Listing}
\title{Listening and Seeing Again: Generative Error Correction for \\ Audio-Visual Speech Recognition}
\author{
    Rui Liu\textsuperscript{\rm 1}\thanks{Corrposending Author.},
    Hongyu Yuan\textsuperscript{\rm 1},
    Haizhou Li\textsuperscript{\rm 2, 3}
}
\affiliations{
    \textsuperscript{\rm 1}Inner Mongolia University, China\\
    \textsuperscript{\rm 2}Shenzhen Research Institute of Big Data, School of Data Science, The Chinese University of Hong Kong, Shenzhen, China\\
    \textsuperscript{\rm 3}Department of Electrical and Computer Engineering, National University of Singapore, Singapore\\
    liurui\_imu@163.com, yuanhongyu\_1997@163.com,  haizhouli@cuhk.edu.cn
}

\usepackage{bibentry}
\begin{document}

\maketitle

\begin{abstract}
Unlike traditional Automatic Speech Recognition (ASR), Audio-Visual Speech Recognition (AVSR) takes audio and visual signals simultaneously to infer the transcription. Recent studies have shown that Large Language Models (LLMs) can be effectively used for Generative Error Correction (GER) in ASR by predicting the best transcription from ASR-generated N-best hypotheses. However, these LLMs lack the ability to simultaneously understand audio and visual, making the GER approach challenging to apply in AVSR. In this work, we propose a novel GER paradigm for AVSR, termed \textbf{AVGER}, that follows the concept of ``listening and seeing again''. Specifically, we first use the powerful AVSR system to read the audio and visual signals to get the N-Best hypotheses, and then use the Q-former-based Multimodal Synchronous Encoder to read the audio and visual information again and convert them into an audio and video compression representation respectively that can be understood by LLM. Afterward, the audio-visual compression representation and the N-Best hypothesis together constitute a Cross-modal Prompt to guide the LLM in producing the best transcription. In addition, we also proposed a Multi-Level Consistency Constraint training criterion, including logits-level, utterance-level and representations-level, to improve the correction accuracy while enhancing the interpretability of audio and visual compression representations. The experimental results on the LRS3 dataset show that our method outperforms current mainstream AVSR systems. The proposed AVGER can reduce the Word Error Rate (WER) by 24\% compared to them. \textcolor[rgb]{0.93,0.0,0.47}{Code and models can be found at: \url{https://github.com/CircleRedRain/AVGER}}.
\end{abstract}

\vspace{-1.5em}
\section{Introduction}

% \textcolor{red}{Para 1: introduce the ASR and AVSR. highlight their difference. then highlight use the LLM to correct the N-Best results is very important and becomes a hot topic.}

Automatic Speech Recognition (ASR) is a technology that enables the conversion of spoken language into text. It primarily relies on audio signals, using acoustic and linguistic models to understand and transcribe speech \cite{whisper}. Unlike ASR, Audio-Visual Speech Recognition (AVSR) is an advanced form of speech recognition that integrates both audio and visual inputs, such as lip movements, to improve the robustness and accuracy of speech recognition. In recent years, Large Language Models (LLMs) have garnered significant interest in research due to their powerful generative capabilities and extensive intrinsic knowledge \cite{llama, llama2, gpt4}, leading to widespread success in various Natural Language Processing (NLP) tasks. A notable application is Generative Error Correction (GER) \cite{hyporadise, robustger, whisper-llama}.

% \textcolor{red}{Para 2: introduce the traditional GER method for ASR, and list some famous works.}

GER employs LLMs to map the correct transcription from the N-best hypotheses decoded by ASR systems, significantly enhancing ASR results and outperforming typical LM rescoring methods \cite{lm}. It has demonstrated great effectiveness in learning the mapping from hypotheses to transcription through parameter-efficient LLM fine-tuning \cite{lora}. Powered by LLMs, rencent work \cite{hyporadise} proposes a GER benchmark for ASR, and they release a HyPoradise dataset that contains over 332K pairs of decoded N-best hypotheses and ground-truth transcription in various ASR domains. However, the GER benchmark lacks specificity on noisy ASR scenarios, which are the most common in real world. To this end, latest work \cite{robustger} extend the GER benchmark to noisy conditions, as well as propose a Robust HyPoradise (RobustHP) dataset with 113K hypotheses-transcription pairs from various ASR corpus in common noisy scenarios. As LLM's ability to handle multimodal tasks is explored, some work has begun to inject acoustic information into LLM to improve GER performance \cite{whisper-llama, listen-again, uadf, mmger}.

% \textcolor{red}{Para 3: However, the traditional GER method for ASR is Not applicable to AVSR, since cannot understand the audio and visual at the same time. Note that the audio and visual are very important for AVSR.}

However, while some GER methods for ASR can be adapted to AVSR \cite{whisper-llama, mmger}, it does not perform optimally in this context, since it cannot understand the audio and visual at the same time. During conversations, humans do not rely solely on auditory cues but also subconsciously observe the speaker's lip movements \cite{msrl}, as visual cues from this region provide valuable information for understanding speech. These visual cues become particularly crucial in noisy environments or when the speaker is speaking softly \cite{msrl}. For AVSR systems that integrate both auditory and visual information, this phenomenon remains critically important. Extensive research has shown that incorporating noise-independent visual cues can significantly enhance the noise robustness and recognition accuracy of speech recognition systems, underscoring the equal importance of visual information and auditory signals \cite{ctc-attention, conformer, avhubert_avsr, auto_avsr}. Therefore, extending GER techniques to incorporate visual information in AVSR systems is a natural progression to improve their effectiveness.

% \textcolor{red}{Para 4: To address this issue, we propose AVGER, a novel GER paradigm for AVSR. Note that AVGER follows the concept of “listening and seeing again”.}

To address the gap in AVSR error correction research, we propose a method named AVGER, which corrects AVSR results using the original audio and video information. Note that AVGER follows the concept of ``listening and seeing again''. Specifically, the original audio signals and lip videos are input into a Multimodal Synchronous Encoder based on the Q-Former \cite{blip2, seg_qformer}, which operates independently from the AVSR system. This setup is designed to extract semantic feature representations specifically tailored for error correction in LLMs. These representations are first combined with the N-best hypotheses through a Cross-modal Prompt, and then input into an LLM. The LLM leverages efficient parameter-tuning techniques to enhance its understanding of multimodal information, ultimately producing a correct transcription based on the original audiovisual features and the N-best list. Additionally, we customize a Multi-Level Consistency Constraint Loss Function that aligns with specific task metrics and considers multimodal information. This loss function comprises three parts: 1) Logits-Level: The traditional Cross-Entropy (CE) Loss, which minimizes the discrepancy between predicted and true distributions, enhancing the model's accuracy; 2) Utterance-Level: A WER component, identical to the task metric, which bridges the gap between training objectives and task benchmarks, thereby effectively boosting model performance; 3) Representation-Level: A Central Moment Discrepancy (CMD) loss \cite{cmd} that minimizes the gap between audio, visual and textual features, enhancing the model’s ability to understand and process multimodal information. The main contributions of this paper are:

\begin{itemize}
    \item We propose AVGER — a paradigm for AVSR generative error correction that addressing the limitations of current GER approaches in AVSR scenarios. To the best of our knowledge, we are the first to introduce a method that integrates both audio and visual information for GER in AVSR systems.
    \item Multimodal Synchronous Encoder and Cross-modal Prompt to build the link between N-best and raw audio-visual information, and using a Multi-Level Consistency Constraint Loss Function to guide the LLM to comprehend audio-visual information and produce transcriptions that most accurately reflect the actual spoken content. 
    \item Experimental results show that our method enhances the current mainstream AVSR system and effectively reduces the WER by 24\%.
\end{itemize}

\vspace{-1em}
\section{Related Work}

\vspace{-0.5em}
\subsection{Q-former and Temporal Synchronisation}

% \textcolor{red}{Para-1: what is Q-former. some works use Q-former to connect something.  [list some works.]  However, this work is different, we use Q-former to 1) 2) 3) .}

Querying Transformer (Q-Former) consists of two transformer submodules that share the same self-attention layers, proposed by BLIP2 \cite{blip2}. A set number of learnable query parameters as input to the transformer. The queries interact with each other through self-attention layers, and interact with features through cross-attention layers. While Q-Former has been primarily proposed for visual information extraction \cite{video_llama, blip2}, it also performs remarkably in extracting auditory features for generic audio understanding \cite{seg_qformer, salmonn, secap}. In addition, various types of modality aligners have been studied, such as the cross-attention mechanism \cite{flamingo}, pre-trained multimodal embeddings \cite{imagebind}, and temporal and spatial pooling \cite{videochatgpt} \textit{etc}. Unlike these approaches, our method slices temporally aligned audio and video into segments of equal length, processes each segment with Q-Former, and then stacks the results to maintain temporal synchronization.

\vspace{-0.8em}
\subsection{Audio-Visual Speech Recognition} 
\vspace{-0.5em}

AVSR has attracted increasing research interest in improving the performance of speech recognition systems by combining speech signals with visual signals that are not affected by acoustic noise. AV-HuBERT \cite{avhubert_pretatin} learn the correspondence of audio and video modalities in a self-supervised manner, which is further augmented in \cite{avhubert_avsr} to improve noise-robustness. Auto-AVSR \cite{auto_avsr} using publicly-available pre-trained ASR models to automatically transcribe unlabelled datasets to increase the training set size. MRSL \cite{msrl} leveraging visual modality-specific representations that carry noise-invariant information from the visual stream to improve the AVSR system by reinforcement learning. In this paper, we select AV-HuBERT as our AVSR system, primarily due to its robust self-supervised learning capabilities. It effectively captures deep audio and visual features from unlabeled data, which significantly enhances its performance in AVSR tasks within noisy environments.

\vspace{-0.8em}
\subsection{ASR Generative Error Correction} 
\vspace{-0.5em}

Benefiting from the powerful generative capabilities of LLMs, recent studies have introduced a GER benchmark \cite{hyporadise} that uses LLMs to generate the ground-truth transcript from the N-best hypotheses list produced by ASR. RobustGER \cite{robustger} extends this benchmark to noisy conditions, achieving significant performance improvements. As the research progressed, a new problem was identified: due to the uncertainty of LLMs generation, the generated transcripts are sometimes grammatically correct but do not match the original speech. In response, ClozeGER \cite{listen-again} formatted GER as a cloze test that eliminates redundancy of input information and combines it with the source speech for error correction. LipGER \cite{lipger} uses lip movements for error correction of ASR. UADF \cite{uadf} utilises token-level uncertainty estimation at the decision stage to dynamically determine the fusion weights to be assigned to each modality at each decoding step, and hence this approach can be used for AVSR. However, due to the limited semantic information contained in the decision-level information, it does not migrate well to AVSR systems that contain additional visual information. In this paper, we construct a GER method that simultaneously understands audio and video through Cross-modal Prompt and Multi-Level Consistency Constraint.

\begin{figure*}[!t]
\vspace{-1em}
        \centering
        \includegraphics[width=1\linewidth]{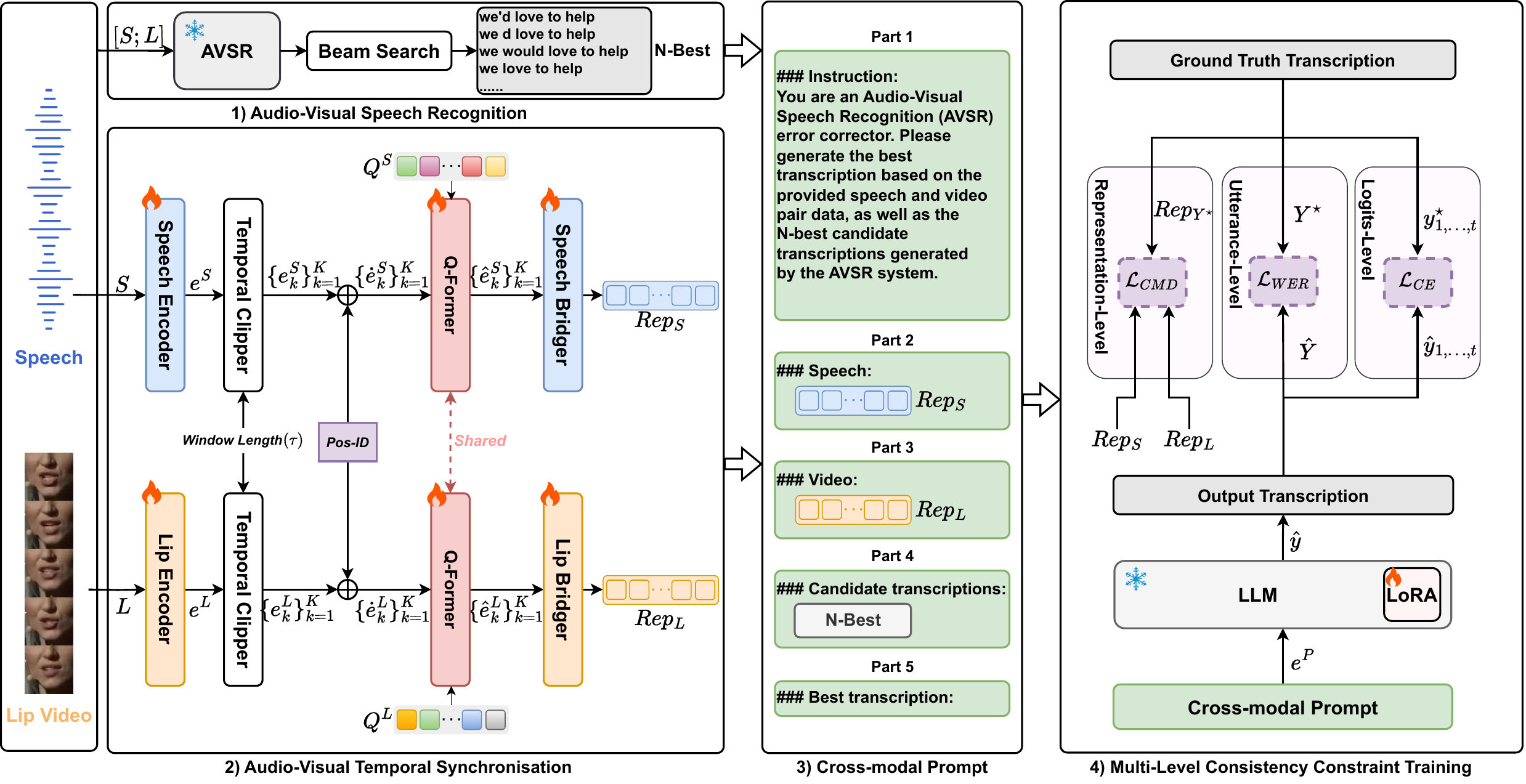}
\vspace{-6mm}
\caption{Architecture of the AVGER System. The AVGER system integrates speech and lip video inputs for improved transcription accuracy. It consists of: 
1) Audio-Visual Speech Recognition;
2) Audio-Visual Temporal Synchronisation;
3) Cross-modal Prompt;
and 4) Multi-Level Consistency Constraint Training.}
        \label{avger}
\vspace{-1.2em}
\end{figure*}

\vspace{-1mm}
\section{Methodology}
% \vspace{-0.5em}
In this section, we will introduce our proposed AVGER. First, we describe the overall architecture in detail. Next, we will elaborate on the principle of Multimodal Synchronous Encoder and Cross-modal Prompt. Finally, we will introduce the Representation-Level Consistency Constraint Loss Function.
\vspace{-0.5em}
\subsection{Main Architecture}
\vspace{-0.3em}
Although the AVSR system has already extracted features from the source speech and lip video, we design an encoder independent of the AVSR to ensure that the LLM does not become coupled with it. This independent encoder re-extracts audio and video representations that are better suited for the LLM. We chose LLaMA \cite{llama} as the foundation for our decoder due to its strong generative capabilities and flexibility in handling complex multimodal inputs, making it particularly well-suited for the nuanced task of AVSR error correction. As shown in Figure 1, AVGER incorporates a Multimodal Synchronous Encoder based on QFormer, a decoder based on LLaMA, and an AVSR system implemented using AV-HuBERT.

The source speech and lip video contain a wealth of information. Initially, the AVSR system processes these inputs and transcribes them into N-best hypotheses $\mathcal{Y}_N = \{Y_1, Y_2, \cdots, Y_N\}$ using beam search decoding. Subsequently, a Q-Former-based Multimodal Synchronous Encoder re-extracts and compresses the temporally aligned frame-level features from the audio and video, capturing subtle linguistic and semantic nuances. Finally, these features, along with the N-best hypotheses, are input into the LLaMA-based decoder to generate the most accurate transcription.

\vspace{-2mm}
\subsection{Multimodal Synchronous Encoder}
% \vspace{-0.3em}
This module is used to extract the representations of the source audio and video signals and compress their length. The module consists of two branches, each containing an Encoder, a Temporal Clipper, a Q-Former shared by both branches, a set of learnable query parameters, and a Bridger.

The temporally aligned speech (denoted as $S$) and lip video (denoted as $L$) pairs with time length $T$ are input into their respective convolution-based encoders to get frame-level features $e^S\in\mathbb{R}^{T_S \times d}$ and $e^L\in\mathbb{R}^{T_L \times d}$, where $T_S$ and $T_L$ are the sequence lengths after processing by Speech Encoder and Lip Encoder, $d$ is the dimensions of hidden state, Note that $T_S$ and $T_L$ are not equal due to differences in the processing of video and audio modalities:
\begin{equation}
\vspace{-0.3em}
\begin{cases}
e^S = \operatorname{SpeechEncoder}(S) \\
e^L = \operatorname{LipEncoder}(L)
\end{cases}
\vspace{-0.3em}
\end{equation}

The Temporal Clipper segments the frame-level features based on the specified time window length $\tau$, ensuring each segment corresponds to a fixed temporal span. Notably, the Temporal Clipper operates without any learnable parameters. Since the LLM can see all temporal windows, there is no overlap in Temporal Clipper. This setting is crucial for aligning the temporal information across modalities, ensuring that the Q-Former to process segments with consistent temporal span from both speech and lip video, thereby ensuring synchronization:
\begin{equation}
\vspace{-0.3em}
\begin{cases}
\{e^S_k\}^K_{k=1} = \operatorname{TemporalClipper}(e^S, \tau) \\
\{e^L_k\}^K_{k=1} = \operatorname{TemporalClipper}(e^L, \tau)
\end{cases}
\vspace{-0.3em}
\end{equation}
where $e^S_k\in\mathbb{R}^{\tau_S \times d}$ and $e^L_k\in\mathbb{R}^{\tau_L \times d}$ denote the $k$-th segment of $e^S$ and $e^L$, with $K=\lceil T / \tau \rceil$, $\tau_S=T_S/K$, and $\tau_L=T_L/K$.

Segment-level position embeddings $\{P_k\}^K_{k=1}, P_k\in\mathbb{R}^d$ are added to each $e^S_k$ and $e^L_k$. The addition of segment-level position embeddings provides the Q-Former with temporal context, ensuring that the sequential order and timing relationships between segments are preserved during processing, which is essential for maintaining the integrity of temporal features:
\begin{equation}
\begin{cases}
\{\dot{e}^S_k\}^K_{k=1} = \{e^S_k \oplus P_k\}^K_{k=1} \\
\{\dot{e}^L_k\}^K_{k=1} = \{e^L_k \oplus P_k\}^K_{k=1}
\end{cases}
\end{equation} 
where $\oplus$ denotes the vector addition operation, $\dot{e}^S_k\in\mathbb{R}^{\tau_S \times d}$, and $\dot{e}^L_k\in\mathbb{R}^{\tau_L \times d}$.

\begin{figure}[!t]
\vspace{-1em}
        \centering
        \includegraphics[width=1\linewidth]{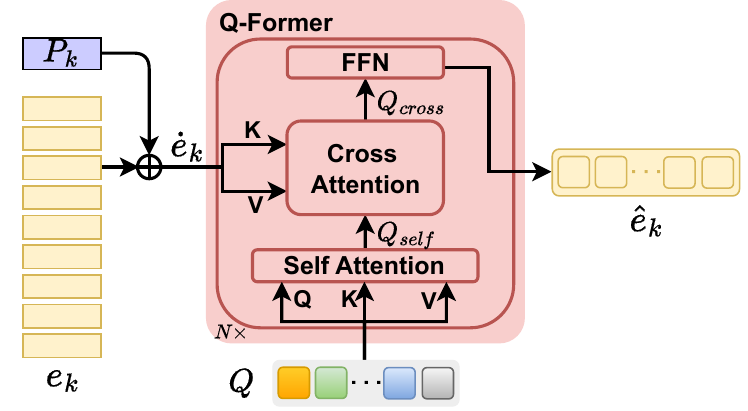}
        \vspace{-3mm}
\caption{The workflow of Q-Former. $e_k$ is a clip of segmented frame-level features, $P_k$ is segment-level position embeddings, and $\oplus$ denotes the vector addition operation.}
\vspace{-5mm}
        \label{qformer}

\end{figure}

A shared Q-Former, with specific learnable query parameters for each modality, is employed to process these segments added segment-level position embeddings $\dot{e}^S_k$ and $\dot{e}^L_k$. These parameters are expressed as $Q^S\in\mathbb{R}^{n_q \times d}$ for speech and $Q^L\in\mathbb{R}^{n_q \times d}$ for lip video, where $n_q$ is the length of learnable query parameters, each with the same length. The use of different learnable query parameters for speech and lip video allows the Q-Former to capture modality-specific features while aligning them in a shared representation space, facilitating effective multimodal fusion and enhancing the model's ability to handle cross-modal information. 

Using $Q\in\mathbb{R}^{n_q \times d}$ as a proxy for $Q_S$ and $Q_L$, the Q-Former is computed as shown in Figure 2. First, input $Q$ into the self-attention mechanism to capture and model the internal dependencies and relationships between all pairs of query vectors. This process allows the Q-Former to dynamically reweight the importance of each query vector based on its relevance to others:
\begin{equation}
    Q_{self} = \operatorname{Softmax}\left(\frac{QW_{q_{self}} \left(QW_{k_{self}}\right)^T}{\sqrt{d_k}}\right) QW_{v_{self}}
\end{equation}
where $W_{q_{self}}$, $W_{k_{self}}$ , and $W_{v_{self}}$ are the learnable weight matrices for queries, keys, and values in the self-attention mechanism. Using $\dot{e}$ as a proxy for $\dot{e}^S_k$ and $\dot{e}^L_k$. Next, the cross-attention mechanism is applied to interact $Q_{self}$ with $\dot{e}$ to effectively integrate and align information across the modalities. This interaction allows the model to selectively focus on and combine the most relevant features from the segment-level representations $\dot{e}$:
\begin{equation}
    Q_{cross} = \operatorname{Softmax}\left(\frac{Q_{self} W_{q_{cro}} \left(\dot{e} W_{k_{cro}}\right)^T}{\sqrt{d_k}}\right) \dot{e} W_{v_{cro}}
\end{equation}
where $W_{q_{cro}}$, $W_{k_{cro}}$ , and $W_{v_{cro}}$ are the learnable weight matrices for queries, keys, and values in the cross-attention mechanism. Finally the output is obtained through the FFN layer: $\hat{e} = \operatorname{FFN}(Q_{cross})$, where $\hat{e}$ is a proxy for $\hat{e}^S_k$ and $\hat{e}^L_k$. The above Q-former calculation process is represented as follows:
\begin{equation}
    \hat{e} = \operatorname{Q\text{-}Former}(Q, \dot{e})
\end{equation}
Based on this, we obtain this equation:
\begin{equation}
\begin{cases}
\{\hat{e}^S_k\}^K_{k=1} = \{\operatorname{Q\text{-}Former}(Q^S, \dot{e}^S_k)\}^K_{k=1} \\
\{\hat{e}^L_k\}^K_{k=1} = \{\operatorname{Q\text{-}Former}(Q^L, \dot{e}^L_k)\}^K_{k=1}
\end{cases}
\end{equation} 

Finally, two bridge modules are utilized to stack the Q-Former-processed representations into two complete sequences and map them to the embedding space of the LLM through a linear layer. This process ensures that the representations from both modalities (speech and lip video) are transformed into a unified format that the LLM can effectively process. In this case, the two modal representations have the same length and are aligned, which is crucial for enabling the LLM to jointly consider information from both modalities:
\begin{equation}
\begin{cases}
Rep_S = \operatorname{SpeechBridger}(\{\hat{e}^S_k\}^K_{k=1}) \\
Rep_L = \operatorname{LipBridger}(\{\hat{e}^L_k\}^K_{k=1})
\end{cases}
\end{equation} 

\vspace{-1mm}
\subsection{Cross-modal Prompt}
% \vspace{-0.5em}
To fully leverage the semantic understanding capabilities of LLMs and generate more accurate transcriptions, we design a structured Cross-modal Prompt. This prompt is intended to guide the model in effectively integrating multimodal information from both speech and video inputs.

The Cross-modal Prompt is structured into five distinct parts, each defined by a unique heading to guide the model’s processing:
\begin{enumerate}[1)]
\item \textit{\#\#\# Instruction}: This part sets the context for the task and instructs the model on what is expected. 
\item \textit{\#\#\# Speech}: It represents where the $Rep_S$ is inputted into the model. 
\item \textit{\#\#\# Video}: It indicates where the $Rep_L$ is inputted. 
\item \textit{\#\#\# Candidate transcriptions}: It indicates the position of the N-best hypotheses list $\mathcal{Y}_N$ obtained from the AVSR system, which the model uses along with the video representation $Rep_L$ and the audio representation $Rep_S$ for transcript correction. 
\item \textit{\#\#\# Best transcription}: The final part is where the model outputs the corrected transcript,
\end{enumerate}

We detail the process of constructing the LLM input embedding. Initially, the N-Best hypotheses list $\mathcal{Y}_N$ is placed into the \textit{\#\#\# Candidate transcriptions} section of the prompt. This prompt, now containing the hypotheses, is tokenized \cite{bpe} and passed through the LLM’s embedding layer to generate text embeddings. Finally, the speech representation $Rep_S$ and the lip video representation $Rep_L$, both preprocessed to match the LLM’s input dimensions, are inserted into their respective parts—\textit{\#\#\# Speech} and \textit{\#\#\# Video}—completing the creation of the LLM input embedding.

\vspace{-1mm}
\subsection{Multi-Level Consistency Constraint Training}
% \vspace{-0.5em}
The goal of GER is to learn a hypotheses-to-transcription (H2T) mapping $\mathcal{M}_{H2T}$ that predicts the transcription $\hat{Y}$ based on N-best list $\mathcal{Y}_N$ :
\begin{equation}
    \hat{Y} = \mathcal{M}_{H2T}(\mathcal{Y}_N)
\end{equation}

Given the ground-truth transcription $Y^\star$ , we can finetune the LLM to learn $\mathcal{M}_{H2T}$ in an auto-regressive manner. To effectively learn the hypotheses-to-transcription mapping $\mathcal{M}_{H2T}$, we propose a Multi-Level Consistency Constraint Loss Function. This loss function comprises three levels: Representation-Level Consistency Constraint, Utterance-Level Consistency Constraint, and Logits-Level Consistency Constraint.

\vspace{-0.5em}
\subsubsection{Representation-Level Consistency Constraint}
CMD based distance constraint aims to reduce the discrepancy between the features of different modalities, so that the semantic features of  $Rep_S$ and $Rep_L$ are closer to those of $Rep_{Y^\star}$. Note that CMD \cite{cmd} is a state-of-the-art distance metric that measures the discrepancy between the distribution of two features by matching their order-wise moment differences. In the training phase, we will process the ground-truth transcription $Y^\star$ through the tokenizer and LLM embedding layer to get the text representation $Rep_{Y^\star}$. We ensure that the extracted $Rep_S$ and $Rep_L$ are closest to the $Rep_{Y^\star}$ by minimising the $\mathcal{L}_{CMD}$:
\begin{equation}
\begin{aligned}
    \mathcal{L}_{CMD} = \frac{1}{3} 
    &\sum_{\begin{subarray}{l}
        (m_1, m_2)\\
         \in \{(S,L), \\
        (S,{Y^\star}),\\
        (L,{Y^\star}) \}
     \end{subarray}}
    (\left\| \mathbb{E}(Rep_{m_1}) - \mathbb{E}(Rep_{m_2}) \right\|_2 \\&
    + \sum_{k=5}^{K} \left\| C_k(Rep_{m_1}) - C_k(Rep_{m_2}) \right\|_2)
\end{aligned}
\end{equation}
where $\mathbb{E}(Rep)$ is the empirical expectation vector of the input sample $Rep$, and $C_k(Rep) = \mathbb{E}((Rep-\mathbb{E}(Rep))^k)$ is the vector of all $k^{th}$ order sample central moments of the coordinates of $Rep$.

\vspace{-0.5em}
\subsubsection{Utterance-Level Consistency Constraint}
The Word Error Rate loss $\mathcal{L}_{WER}$ is designed to minimize the number of insertions, deletions, and substitutions needed to convert the transcription $\hat{Y}$ into the ground-truth transcription $Y^\star$ \cite{wer}. By focusing on these specific types of errors, it directly improves the intelligibility and accuracy of the generated text at the utterance-level: 
\begin{equation}
    \mathcal{L}_{WER} = \frac{D(\hat{Y}||Y^\star)}{\operatorname{len}(Y^\star)} 
\end{equation}
where $D(\cdot||\cdot)$ denotes the Levenshtein Distance between two sequence, and $\operatorname{len}(\cdot)$ denotes the number of words in sequence. 

\vspace{-0.5em}
\subsubsection{Logits-Level Consistency Constraint}
The Cross-Entropy loss $\mathcal{L}_{CE}$ \cite{attention} is widely used in classification tasks to measure the discrepancy between the predicted probability distribution and the ground-truth distribution. By penalizing the model for inaccurate predictions, it encourages the model to generate probabilities that are closer to the true distribution:
\begin{equation}
    \mathcal{L}_{CE} = \sum_{t=1}^{T} -\log P_{\theta}(y_t^* \mid y_{t-1}^*, \ldots, y_1^*, \mathcal{Y}_N)
\end{equation}
where $y_t^*$ is the $t$-th token of $Y^\star$ , and $\theta$ denotes the learnable parameters in LLM (\textit{i}.\textit{e}., LoRA \cite{lora}). 

The Multi-Level Consistency Constraint Loss $\mathcal{L}_{MLC}$ combines the three loss components to provide a comprehensive training signal, ensuring consistency at the representation-level, utterance-level, and logits-level:
\begin{equation}
    \mathcal{L}_{MLC} = \mathcal{L}_{CMD} + \mathcal{L}_{WER} + \mathcal{L}_{CE}
\end{equation}

LoRA (Low-Rank Adaptation) is a technique designed for efficiently fine-tuning LLMs. Typically, fine-tuning LLMs requires updating a substantial number of parameters, which can be computationally expensive and prone to overfitting. LoRA addresses this challenge by employing low-rank decomposition, breaking down the weight matrices in the model into the product of lower-rank matrices. This approach significantly reduces the number of parameters that need to be updated during fine-tuning. By doing so, LoRA not only maintains the model's performance but also greatly reduces the computational resources and memory required for fine-tuning. Therefore, we chose LoRA to fine-tuning the LLM to learn $\mathcal{M}_{H2T}$.

\begin{table*}[!htb]		
\vspace{-1em}						
\centering								
\begin{tabular}{@{}cccccccc@{}}								
\toprule								
&		& \multicolumn{2}{c}{GER}		& \multicolumn{2}{c}{UADF}		& \multicolumn{2}{c}{AVGER}		\\ \cmidrule(l){3-8} 
\multirow{-2}{*}{\begin{tabular}[c]{@{}c@{}}SNR\\ (\textit{Babble})\end{tabular}}	& \multirow{-2}{*}{\begin{tabular}[c]{@{}c@{}}AV-HuBERT\\ $\operatorname{WER}_{(\%)}(\downarrow)$\end{tabular}}	& \multicolumn{1}{l}{$\operatorname{WER}_{(\%)}(\downarrow)$}	& \multicolumn{1}{l|}{$\operatorname{WERR}_{(\%)}(\uparrow)$}	& \multicolumn{1}{l}{$\operatorname{WER}_{(\%)}(\downarrow)$}	& \multicolumn{1}{l|}{$\operatorname{WERR}_{(\%)}(\uparrow)$}	& \multicolumn{1}{l}{$\operatorname{WER}_{(\%)}(\downarrow)$}	& \multicolumn{1}{l}{$\operatorname{WERR}_{(\%)}(\uparrow)$}	\\ \midrule
\multicolumn{1}{c|}{-10}	& \multicolumn{1}{c|}{30.3}	& 33.6	& {-10.2} 	& \multicolumn{1}{|c}{\textbf{21.8}}	& {\textbf{28.1}}	& \multicolumn{1}{|c}{29.1}	& {4.0} 	\\
\multicolumn{1}{c|}{-5}	& \multicolumn{1}{c|}{13.5}	& 13.6	& {1.4}  	& \multicolumn{1}{|c}{\textbf{10.7}}	& {\textbf{20.7}}	& \multicolumn{1}{|c}{12.3}	& {8.9} 	\\ \midrule
\multicolumn{1}{c|}{0}	& \multicolumn{1}{c|}{4.9}	& 4.4	& {8.3} 	& \multicolumn{1}{|c}{4.6}        	& {6.1}	& \multicolumn{1}{|c}{\textbf{3.7}}	& {\textbf{24.5}}	\\
\multicolumn{1}{c|}{5}	& \multicolumn{1}{c|}{2.5}	& 2.3	& {4.2} 	& \multicolumn{1}{|c}{2.6}       	& {-4.0}	& \multicolumn{1}{|c}{\textbf{1.9}}	& {\textbf{24.0}}	\\ \midrule
clean 	& 1.45                      	& 1.31 	& {9.66}	& 1.36	& {6.2}	& \textbf{1.10}	& {\textbf{24.1}}	\\ \bottomrule
\end{tabular}				 \vspace{-2mm}	
\caption{$\operatorname{WER}_{(\%)}$ and $\operatorname{WERR}_{(\%)}$ results on LRS-3, including clean audio and four signal-to-noise ratio (SNR) levels of noise audio. \textit{Babble} is the noise drawn from \cite{musan}. GER is a method proposed by Hyporadise \cite{hyporadise} which corrects errors ONLY through the N-Best list. UADF is the approach proposed by \cite{uadf} that corrects errors using decision-level information fusion. }	
\vspace{-5mm}
\end{table*}

\vspace{-0.8em}
\section{Experiment And Results}
\vspace{-1mm}
\subsection{Dataset}
% \vspace{-0.5em}

We conduct the experiments on LRS3 \cite{lrs3}, which is the largest publicly available dataset for audio-visual speech recognition task. LRS3 consists of 151 819 video clips from TED talks with a total of 439 hours. It contains 118 516 (408 hours), 31 982 (30 hours) and 1 321 clips (0.9 hours) in the pretraining, training-validation, and test set. The LRS3 dataset is the largest and most widely accepted benchmark for AVSR, and since our method is based on AV-HuBERT, which itself was primarily evaluated on LRS3, we used this dataset in our experiments. Furthermore, we tested our model's generalization by using LRS3 checkpoints on other datasets.

To train our model within our resource limitations, we combined the original training and test sets from LRS3. After removing data segments longer than 20 seconds, we re-divided the dataset into a new 270h training set, and a 1h validation set, and maintained the original 0.9h test set. We construct the noisy data on the clean LRS3 using \textit{Babble} noise which is drawn from \cite{musan} at four SNR levels: \{-10, -5, 0, 5\}dB, with a 1:1 ratio of noisy to clean data. This configuration is consistent with UADF \cite{uadf} for ease of comparison. %\textcolor{blue}{To the best of our knowledge, babble noise is among the most challenging for speech recognition tasks. Although our method is only trained on babble, it still has strong generalization to other noise types, even unseen ones.}

For each video clip, we detect the 68 facial keypoints using dlib \cite{dlib} and align each frame to a reference face frame via affine transformation. We crop a $96 \times 96$ region-of-interest (ROI) centered on the mouth.

\vspace{-1mm}
\subsection{Experimental Setup}
% \vspace{-0.5em}
We employ LLaMA-7B \cite{llama} as the foundation model and AV-HuBERT as the AVSR system for generating N-best hypotheses. A low-rank adapter is inserted into each layer of LLaMA with the rank of 32 for fine-tuning. 

In the Multimodal Synchronous Encoder, the speech encoder is implemented using the first layer of HuBERT \cite{hubert}, and the lip encoder is derived from the first layer of VideoMAE \cite{videomae}, both leveraging convolutional layers. 
For the lip encoder, we modify the original model to enable the processing of video inputs. We divide each frame of the video clip into four $48 \times 48$ patches. The shared parameters Q-Former comprises 6 Q-Former Blocks \cite{blip2}, with learnable query parameters of length 20. The time window length is set to 1 second. We use a uniform Cross-modal Prompt template to transform the 10-best candidate transcripts and audio-visual representations into inputs suitable for LLMs. 

During fine-tuning, we employ AdamW \cite{adamw} optimizer with a learning rate set to $1\mathrm{e}^{-4}$ and warmup steps set to 0.05\%. The number of training epochs is set to 2, the batch size is set to 256 by using gradient accumulation. The maximum input sequence length is set to 2048. 
% For inference, we adopt top-k and top-p sampling strategies at the same time, where k = 2 and p = 0.8. The temperature is set to 0.2.

To validate the effectiveness of our proposed AVGER system, we conducted comparisons against baseline models such as GER \cite{hyporadise} and UADF \cite{uadf}. We also performed ablation studies to assess the contribution of each module within AVGER. Additionally, we evaluated the impact of various hyperparameter settings on the final performance. The results of these experiments are detailed in the following section.

\vspace{-0.8em}
\subsection{Metrics}
\vspace{-0.2em}
We evaluate performance using WER \cite{wer}, and Word Error Rate Reduction (WERR) \cite{fastCorrect}. Lower values of WER indicate better performance, while a higher WERR signifies a greater improvement relative to the GER approach. 

\vspace{-0.8em}
\subsection{Baselines}
\vspace{-0.2em}
\subsubsection{GER} 
GER is a method proposed by Hyporadise \cite{hyporadise}. We follow the hypotheses-to-transcription-low-rank-adaptation (H2T-LoRA) pipeline in Hyporadise, which avoids tuning the entire set of parameters of the pre-trained model by inserting a neural module with a small number of additional trainable parameters to approximate the full set of parameters, thus learning the H2T mappings efficiently without affecting the LLM pre-training parameters, thus avoiding the need to tune the entire set of parameters of the pre-trained model, and thus learning the H2T mapping efficiently without affecting the LLM pre-training parameters. This pipeline corrects errors ONLY through the N-Best list.

\vspace{-0.5em}
\subsubsection{UADF}
UADF is Uncertainty Aware Dynamic Fusion, an ASR error correction approach proposed by \cite{uadf}. UADF is a multimodal fusion approach implemented into an auto-regressive decoding process and works in two stages: 1) It first analyzes and calibrates the token-level LLM decision, and 2) it then dynamically assimilates the information from the acoustic modality. Since it only relies on decision-level information, it can be directly used in AVSR error correction.

\begin{table*}[]		
\vspace{-1em}							
\centering									
\begin{tabular}{@{}c|c|ccccc|cc@{}}									
\toprule									
\multirow{2}{*}{Model}	& \multirow{2}{*}{ID}	& \multicolumn{5}{c|}{Modules}					& \multicolumn{2}{c}{Metrics} 		\\
	&	& Segment	& Speech	& Lip Video	& $\mathcal{L}_{CMD}$	& $\mathcal{L}_{WER}$	& $\operatorname{WER}_{(\%)}(\downarrow)$	& $\operatorname{WERR}_{(\%)}(\uparrow)$	\\ \midrule
AV-HuBERT               	& \#0	& -      	& -      	& -        	& -   	& -    	& 1.45	& 0	\\ \midrule
\multirow{7}{*}{AVGER} 	& \#1	& \Checkmark  	& \Checkmark  	& \Checkmark  	& \Checkmark  	& \Checkmark   	& 1.10	& 24.1	\\
                        	& \#2	& \Checkmark  	& \Checkmark  	& \Checkmark  	& \Checkmark  	& \XSolidBrush 	& 1.25	& 13.7	\\
                        	& \#3	& \Checkmark  	& \Checkmark  	& \Checkmark  	& \XSolidBrush	& \Checkmark   	& 1.26	& 13.1	\\
                        	& \#4	& \Checkmark  	& \Checkmark  	& \Checkmark  	& \XSolidBrush	& \XSolidBrush 	& 1.30	& 10.3	\\
                        	& \#5	& \Checkmark  	& \Checkmark  	& \XSolidBrush	& \Checkmark  	& \Checkmark   	& 1.19	& 17.9	\\
                        	& \#6	& \Checkmark  	& \XSolidBrush	& \Checkmark  	& \Checkmark  	& \Checkmark   	& 1.27	& 12.4	\\
                        	& \#7	& \XSolidBrush	& \Checkmark  	& \Checkmark  	& \Checkmark  	& \Checkmark   	& 1.31	& 9.6	\\ \bottomrule
\end{tabular}									
\caption{Results of ablation experiments on clean audio. We conducted ablation experiments on AVGER in five areas: 1) Segment: whether or not to slice the frame-level features $e^S$ and $e^L$; 2) Speech: whether or not to process speech; 3) Lip Video: whether or not to process lip video; 4) $\mathcal{L}_{CMD}$: whether or not to compute the CMD loss; 5) $\mathcal{L}_{WER}$: whether or not to calculate WER loss.}	
\vspace{-1.2em}
\end{table*}

\begin{table}[]		
% \vspace{-0.5em}		
\centering				
\begin{tabular}{@{}c|c|cc@{}}				
\toprule				
\multirow{2}{*}{HyperParams}	& \multirow{2}{*}{Value} 	& \multicolumn{2}{c}{Metrics}		\\
	&    	& $\operatorname{WER}_{(\%)}(\downarrow)$	& $\operatorname{WERR}_{(\%)}(\uparrow)$ 	\\ \midrule
\multirow{3}{*}{N-Best}	& 5 	&    1.17	&    19.3	\\
	& 8	&    1.13	&    22.1	\\
	& 10	&    1.10	&    24.1	\\ \midrule
\multirow{3}{*}{Q-Tokens}	& 10	&    1.16	&    20.0	\\
	& 15 	&    1.12	&    22.8	\\
	& 20 	&    1.10	&    24.1	\\ \bottomrule
\end{tabular}				
\caption{Results of HyperParams experiments on clean audio. We investigate the impact of the number of N-Best hypotheses inserted in the Cross-modal prompt and the length of the learnable query parameters on the final error correction results.}		
\vspace{-1.2em}		
\end{table}

\vspace{-0.8em}
\subsection{Results and Analysis}
\vspace{-0.2em}
In this section, we conduct experiments to address the following questions: (i) How does the GER method, which performs well in ASR, translate to AVSR in terms of performance? (ii) Is the proposed AVGER method effective and does it outperform the baseline? (iii) What is the contribution of the modular design in AVGER to the error correction process? (iv) How do different hyperparameter settings affect the final result?

\vspace{-0.5em}
\subsubsection{Main Results} We compared the performance of different models for error correction of AV-HuBERT results in band noise frequency and clean audio conditions, including the WER and WERR of GER, UADF, and the proposed AVGER method. As shown in Table 1, AVGER performs well in most of the SNR conditions, with an average WER of 11.8\%, and positively improves all SNR levels. The WER of AVGER was lowest in the clean speech condition at 1.10\%. In addition, AVGER's WERR reaches 24.5\% and 24.0\% at 0dB and 5dB SNR levels, respectively, significantly outperforming the other methods. This demonstrates the exciting robustness and superior performance of AVGER in dealing with multiple noise environments.

% \textcolor{blue}{
We find that the error correction effect is inconsistent at high SNR (0dB, 5dB) and low SNR (-5dB, -10dB), and the same phenomenon also exists in UADF. We guess the reasons are as follows. At high SNR, the audio signal and the N-Best list are used as the main basis for error correction. At low SNR, the semantic information contained in the audio signal is masked by the noise, and the semantic information contained in the low-quality N-Best generated by the low-quality speech is scarce, so the error correction relies more on the noise-invariant lip video signal. Due to the complexity of video modality processing, our existing training data is insufficient for fine-tuning the LLaMA of plain text to be able to understand audio and video well at the same time. However, this phenomenon is not the focus of this paper. We will try to solve this problem by using multimodal LLMs and increasing the training data in the future.

\vspace{-0.5em}
\subsubsection{Ablation Results} The results of the ablation experiments in Table 2 show that the AVGER performs best when the modules work in concert. Removing the $\mathcal{L}_{WER}$ and $\mathcal{L}_{CMD}$ significantly degrades the performance, suggesting that our Multi-Level Consistency Constraint loss function can effectively guide LLM to understand multimodal information and learn the $\mathcal{M}_{H2T}$. In addition, removing either of the speech and lip videos resulted in varying degrees of increase in WER, showing that each modality acts as an effective gain to the final result. It is worth noting that the performance degradation is most obvious when the frame-level features are not sliced, which implies that Q-Former is difficult to effectively capture the key information when dealing with long sequences, suggesting that our method is able to better maintain the consistency of the temporal information and enhance the inter-modal synchronization by slicing the frame-level features, which in turn improves the overall performance of error correction.

\vspace{-0.5em}
\subsubsection{Further Analysis} Table 3 shows the effects of the number of N-Best hypotheses and the length of learnable query parameters on the final error correction results. As the number of N-Best hypotheses increases from 5 to 10, the WER gradually decreases, which indicates that increasing the number of hypotheses can significantly improve the error correction performance of the model. Similarly, as the number of Q-Tokens increases from 10 to 20, the error correction performance of the model gradually improves, which indicates that longer query parameters can capture more semantic information.

% \vspace{-0.8em}
% \subsection{\textcolor{blue}{Case study}}
% xxx

\vspace{-1.5em}
\section{Conclusion}
\vspace{-0.5em}
In this paper, we present AVGER, a new error correction framework that fills the research gap of GER on AVSR systems. AVGER achieves multimodal alignment by segmenting frame-level features and processing each segment by Q-Former, and guides LLM learning of H2T mappings via a Multi-Level Consistency Constraint loss function. Experimental results show that our AVGER achieves better WER performance improvement on the LRS3 dataset under both clean and noisy conditions, and further analyze the reasons for the performance differences under different SNR configurations and the effectiveness of different modules in the framework.

\vspace{-1em}
\section{Limitations And Future Work}
\vspace{-0.5em}
Although the AVGER demonstrates excellent performance in noisy environments and significantly improves the accuracy of audio-visual speech recognition, it has some limitations that cannot be overlooked. First, the model is primarily trained on the LRS3 dataset, which lacks diversity in speech scenarios and languages, potentially limiting its generalization capabilities. Moreover, the model's robustness against extreme or previously unseen noise types still requires further improvement. Future work will focus on expanding the dataset, enhancing the noise handling capabilities, and exploring the applicability of the system in a variety of speech environments in order to increase the versatility and utility of the AVGER system.

\bibliography{aaai25}

\clearpage

\onecolumn
\section{Appendix. Case Study}
\begin{table}[htbp]
\centering
\begin{tabular}{@{}lll@{}}
\toprule
\multicolumn{1}{l|}{ID}           & \multicolumn{1}{l|}{\#1}                                                                       & \#2                                                                                            \\ \midrule
\multicolumn{1}{l|}{SNR}         & \multicolumn{1}{l|}{Clean}                                                  & Clean                                             \\ 
\multicolumn{1}{l|}{Ground Truth} & \multicolumn{1}{l|}{efficiency is for robots}                                                  & we have problems that we desperately need to solve                                             \\ 
\multicolumn{1}{l|}{AV-HuBERT}    & \multicolumn{1}{l|}{efficiency is \textcolor{red}{4} robots}                                  & \textcolor{red}{i mean} we have problems \textcolor{red}{\_} we desperately need to solve     \\ 
\multicolumn{1}{l|}{GER}          & \multicolumn{1}{l|}{efficiency is \textcolor{green}{for} robots}                              & \textcolor{red}{i mean} we have problems \textcolor{green}{that} we desperately need to solve \\ 
\multicolumn{1}{l|}{UADF}         & \multicolumn{1}{l|}{efficiency is \textcolor{green}{for} robots}                              & we have problems \textcolor{red}{\_} we desperately need to solve                              \\ 
\multicolumn{1}{l|}{AVGER}        & \multicolumn{1}{l|}{efficiency is \textcolor{green}{for} robots}                              & we have problems \textcolor{green}{that} we desperately need to solve                          \\ \midrule
                                  %&                                                                                               &                                                                      \\ \midrule
\multicolumn{1}{l|}{ID}           & \multicolumn{1}{l|}{\#3}                                                                       & \#4                                                                          \\ \midrule
\multicolumn{1}{l|}{SNR}         & \multicolumn{1}{l|}{-10dB}                                                  & -10dB                                             \\ 
\multicolumn{1}{l|}{Ground Truth} & \multicolumn{1}{l|}{not a single one}                                                          & they're just waiting for their day in court                                    \\ 
\multicolumn{1}{l|}{AV-HuBERT}    & \multicolumn{1}{l|}{\textcolor{red}{that is} a single one}                                    & they're just waiting for their \textcolor{red}{data corn}                     \\ 
\multicolumn{1}{l|}{GER}          & \multicolumn{1}{l|}{\textcolor{red}{that is} \textcolor{red}{\_} single one}                 & they're just waiting for their \textcolor{red}{data corn}                     \\ 
\multicolumn{1}{l|}{UADF}         & \multicolumn{1}{l|}{\textcolor{green}{not} a single one}                                      & they're just waiting for their \textcolor{green}{day in court}                                                \\ 
\multicolumn{1}{l|}{AVGER}        & \multicolumn{1}{l|}{\textcolor{green}{not} a single one}                                      & they're just waiting for their \textcolor{green}{day} \textcolor{red}{to} \textcolor{green}{court}         \\ \midrule
                                  %&                                                                                               &                                                                      \\ \midrule
\multicolumn{1}{l|}{ID}           & \multicolumn{1}{l|}{\#5}                                                                      & \#6                                                                          \\ \midrule
\multicolumn{1}{l|}{SNR}         & \multicolumn{1}{l|}{-5dB}                                                  & -5dB                                             \\ 
\multicolumn{1}{l|}{Ground Truth} & \multicolumn{1}{l|}{you might get impeached for that}                                         & we have no suspects                                                          \\ 
\multicolumn{1}{l|}{AV-HuBERT}    & \multicolumn{1}{l|}{you might get \textcolor{red}{impeced} for that}                         & we have no \textcolor{red}{sess backs}                                       \\ 
\multicolumn{1}{l|}{GER}          & \multicolumn{1}{l|}{you might get \textcolor{red}{impeced} for that}                         & we have no \textcolor{red}{seen backs}                                      \\ 
\multicolumn{1}{l|}{UADF}         & \multicolumn{1}{l|}{you might get \textcolor{green}{impeached} for that}                     & we have no \textcolor{green}{suspects}                                      \\ 
\multicolumn{1}{l|}{AVGER}        & \multicolumn{1}{l|}{you might get \textcolor{green}{impeached} for that}                     & we have no \textcolor{green}{suspects}                                       \\ \midrule
                                  %&                                                                                               &                                                                              \\ \midrule
\multicolumn{1}{l|}{ID}           & \multicolumn{1}{l|}{\#7}                                                                      & \#8                                                                          \\ \midrule
\multicolumn{1}{l|}{SNR}         & \multicolumn{1}{l|}{0dB}                                                  & 0dB                                             \\ 
\multicolumn{1}{l|}{Ground Truth} & \multicolumn{1}{l|}{and he said on the spot absolutely yes}                                   & you do not do much editing                                                 \\ 
\multicolumn{1}{l|}{AV-HuBERT}    & \multicolumn{1}{l|}{\textcolor{red}{\_} he said on the spot absolutely yes}                   & you do not do much \textcolor{red}{shell it in}                                     \\ 
\multicolumn{1}{l|}{GER}          & \multicolumn{1}{l|}{\textcolor{green}{and} he said on \textcolor{red}{a} spot absolutely }  & you do not do much \textcolor{red}{shearing}                                 \\ 
\multicolumn{1}{l|}{UADF}         & \multicolumn{1}{l|}{\textcolor{green}{and} he said on the spot absolutely yes}               & you do not do much \textcolor{red}{shedding}                              \\ 
\multicolumn{1}{l|}{AVGER}        & \multicolumn{1}{l|}{\textcolor{green}{and} he said on the spot absolutely yes}                & you do not do much \textcolor{green}{editing}                                   \\ \midrule
                                  %&                                                                                               &                                                                      \\ \midrule
\multicolumn{1}{l|}{ID}           & \multicolumn{1}{l|}{\#9}                                                                      & \#10                                                                         \\ \midrule
\multicolumn{1}{l|}{SNR}         & \multicolumn{1}{l|}{5dB}                                                  & 5dB                                             \\ 
\multicolumn{1}{l|}{Ground Truth} & \multicolumn{1}{l|}{it's the best way to shut people down on an airplane}                     & has it gotten better                                                         \\ 
\multicolumn{1}{l|}{AVSR}         & \multicolumn{1}{l|}{it's the best way to shut people down on \textcolor{red}{\_} airplane}    & \textcolor{red}{is} it gotten better                                         \\ 
\multicolumn{1}{l|}{GER}          & \multicolumn{1}{l|}{it's the best way to shut people down on \textcolor{red}{\_} airplane}   &  \textcolor{green}{has} it gotten better                                       \\ 
\multicolumn{1}{l|}{UADF}         & \multicolumn{1}{l|}{it's the best way to shut people down on \textcolor{red}{\_} airplane}   &  \textcolor{red}{is} it gotten better                                    \\ 
\multicolumn{1}{l|}{AVGER}        & \multicolumn{1}{l|}{it's the best way to shut people down on \textcolor{green}{an} airplane}  & \textcolor{green}{has} it gotten better                                      \\ \bottomrule
\end{tabular}
\caption{Case study comparisons of AVSR transcriptions under various noise conditions (SNR) and correction methods. The table presents the ground truth along with outputs from AV-HuBERT, GER, UADF, and AVGER across different SNR levels (Clean, -10dB, -5dB, 0dB, and 5dB). Correct transcriptions are highlighted in green, and errors are marked in red. \textcolor[rgb]{0.93,0.0,0.47}{Demo videos can be found at: \url{https://github.com/CircleRedRain/AVGER}}.}	
\end{table}

\end{document}